\documentstyle{coralgables}
\input boxedeps.tex
\SetRokickiEPSFSpecial  
\HideDisplacementBoxes

\newcommand{\three}{\put(0,-.05){\makebox[0in]{$\times$}}
                    \put(1,-.05){\makebox[0in]{$\times$}}
                    \put(2,-.05){\makebox[0in]{$\times$}}}
\newcommand{\link}{\rule[0.15em]{25mm}{0.5mm}}		
\newcommand{\dd}{\dagger}
\newcommand{\be}{\begin{equation}}
\newcommand{\eq}{\end{equation}}
\newcommand{\Tr}{{\rm \, Tr \!}}

\begin{document}

\vspace*{9\baselineskip}

\noindent
{\bf THE TRANSVERSE LATTICE IN 2+1 DIMENSIONS\footnote{
Presented by B. van de Sande}}\\
\vspace{2\baselineskip}

\noindent\hspace{1in}\begin{tabular}{@{}r@{}l@{}}
&Brett van de Sande,$^{a}$ 
and Simon Dalley$^{b}$\\ \\
$^{a}$&Max Planck Institut f\"ur Kernphysik\\
&Postfach 10.39.80, D-69029 Heidelberg, Germany\\ \\
$^{b}$&Department of Applied Mathematics and Theoretical Physics\\
&Silver Street, Cambridge CB3 9EW, England
\end{tabular}\\

\section*{INTRODUCTION}

Based on an idea due to Bardeen and Pearson, we formulate the
light-front Hamiltonian problem for $SU(N)$ Yang-Mills theory
in $(2+1)$-dimensions using two continuous space-time 
dimensions with the remaining space dimension discretized on a lattice.
We employ analytic and numerical methods to investigate 
the string tension and the glueball spectrum in the 
$N \to \infty$ limit. 
Our preliminary results show qualitative agreement with recent 
Euclidean lattice Monte Carlo simulations.
In the following, we attempt to give a more pedagogical introduction
to the idea of the transverse lattice; more detail
may be found in Ref.~\cite{incest}.

\section*{MOTIVATION}

There exists a large gap between
the quantum field theory of QCD with its many successes in the 
context of perturbation theory and experimental
observables associated with QCD bound states:
the hadron mass spectrum, structure functions, form factors, {\em et cetera}.
There has been little progress during the last 20 years in building 
a bridge between the field theory and these non-perturbative
properties of the hadron spectrum.

Let us contrast two attempts to bridge this gap: 
the Euclidean lattice Monte Carlo (ELMC) approach
and light-front field theory
where one uses Hamiltonian techniqes on a 
theory which is quantized on a surface of constant $x^+ = (x^0+x^3)/\sqrt 2$.
As far as progress is concerned, ELMC is much further along, benefiting
from a large research effort since the mid 1970's. In contrast,  
the light-front approach
is not so far along; most research effort has occured since late 1980's.
For ELMC, further progress is limited mainly by the speed of 
available computers.  For light-front, progress is currently 
limited by conceptual issues, {\em exempli grati} renormalization.
Even when successful, ELMC is not able to measure many interesting 
observables, such as structure functions, directly.
For light-front field theory, physically interesting observables are quite
easily calculated from the bound state wavefunctions.
The approach that we will discuss here, the transverse lattice, 
uses ideas from both the light-front and lattice approaches.

Instead of solving the full theory of QCD, 
we will focus on a particular model:
$SU(N)$ Yang Mills theory in 2+1 dimensions in the $N \to \infty$
limit.  Why 2+1 dimensions instead of 3+1 dimensions? 
Aside from the obvious fact that it has fewer degrees of freedom,
one should note that the theory is super-renormalizable.  In
the context of lattice calculations, this means that there is
no critical point and the associated `critical slowing' is
absent.  Consequently, excellent lattice spectra are available 
for 2+1 dimensions~\cite{teper} which one can use as a comparison.
Why large $N$\/?  The $N \to \infty$ limit allows considerable 
simplification of our computational problem; in addition, the lattice data 
indicates that $1/N$ corrections to the low energy spectrum
are quite small.

\section*{THE TRANSVERSE LATTICE}
A number of years ago Bardeen and Pearson \cite{bard1,bard2} 
formulated a light-front Hamiltonian version of lattice gauge theory,
which  makes use of the fact that two components of the gauge field are 
unphysical. 
In this approach {\em two} spacetime dimensions are kept continuous 
$x^\pm = (x^0\pm x^2)/\sqrt2$ while the remaining transverse 
spatial dimension $x^1$ is discretized on a lattice. 
Lattice sites are labeled by integer-valued index $i$ and the lattice
constant is $a$. 
Following Bardeen and Pearson, we associate longitudinal gauge 
fields $A^\pm_i(x^+,x^-)$ with
lattice site $i$ and link variable $U_i(x^+,x^-)$ with the link
between sites $i$ and $i+1$,
\begin{equation}
\begin{array}{*{5}{@{}c}}
   &\makebox[0in]{$A^\pm_{i}$} &   U_i &
              \makebox[0in]{$A^\pm_{i+1}$} &U_{i+1}      \\
         \link &\times & \link & \times &\link\\
   & \makebox[0in]{$i$}&&  \makebox[0in]{$i+1$}&
    \end{array} \; .
\end{equation}
Note that $A^\pm_i$ and $U_i$ are $N\times N$ color matrices
where $A^\pm_i$ is Hermitian and traceless and $U_i \in SU(N)$.  
We can define the color Maxwell tensor
\begin{equation}
        F^{\alpha,\beta}_i = \partial^\alpha A^\beta_i -
	           \partial^\beta A^\alpha_i + i [A^\alpha,A^\beta]
\end{equation}
where $\alpha,\beta \in \{+,-\}$ 
along with a covariant derivative
\begin{equation}
     D^\alpha U_i = \left(\partial^\alpha +i A^{\alpha}_{i-1}\right)
                    U_i - i U_i A^{\alpha}_{i+1} \; .
\end{equation}
Using this notation, one can write down an action,
\be
A = a \sum_i \int dx^+\, dx^- \, \Tr\left\{-{1 \over 4g^2} 
   F_{\alpha\beta,i} F^{\alpha\beta}_i
+ {1 \over 2 a^2 g^2} D_{\alpha} U_i D^{\alpha} 
U_i^{\dd}\right\} \; , \label{lag}
\eq
which has several important properties:
\begin{itemize}
 \item If we take the naive continuum limit $a \to 0$ where
       $U_i = \exp(-i a A^1_i)$, we recover the continuum 
       Yang-Mills action.
  \item By construction, this action is 100\% gauge invariant.
  \item It has an automatic confinement mechanism. We 
        will say more about this later.
\end{itemize}

\subsection*{Linearization}

In an ideal world, Eqn.~\ref{lag} is the action one would quantize.
However, we are faced with a problem.  The field theory on each link of 
this transverse lattice is an $SU(N)\times SU(N)$ non-linear $\sigma$-model.  
Upon quantizing this theory, we must enforce $N^2$-nonlinear constraints
on each link.  Although some attempt has been made to do this 
quantization~\cite{griff}, success has remained elusive.
Instead, we perform the linearization suggested by the original 
authors~\cite{bard1}.  In this scheme
one replaces the unitary link variables $U_i$ with $N \times N$ complex
matrices\footnote{The $O(n)$ $\sigma$-models 
are a good example of where this linearization is known to work.} 
$M_i$, that is, $U_i \to \sqrt{2ag^2} M_i$.

The obvious next step would be to include an effective potential 
to enforce the constraint $\sqrt{2ag^2} M_i \in SU(N)$ dynamically.
For instance the term\footnote{For large $N$,  we can
ignore the distinction between $U(N)$ and $SU(N)$.}
\begin{eqnarray}
    & &\lambda_c \Tr\left\{\left(M^\dd_i M_i - 2 a g^2\right)^2\right\}
    \label{lac}\\
    &=&   \lambda_c \Tr\left\{M^\dd_i M_i M^\dd_i M_i\right\}
      - 4 a g^2 \Tr\left\{M^\dd_i M_i \right\} + \left(2 a g^2\right)^2 
     \label{lac2}
\end{eqnarray}
is minimized precisely when $\sqrt{2ag^2} M_i \in U(N)$.
One could imagine adding such a term to the Hamiltonian,
taking the $\lambda_c \to \infty$ limit, and recovering the
transverse lattice action (\ref{lag}).  
Unfortunately, a closer inspection reveals  several problems:
\begin{itemize}
\item The second term in (\ref{lac2}) is a mass term with negative a 
$\left(\mbox{mass}\right)^2$ coefficient.  Quantization of the theory 
breaks down for negative $\left(\mbox{mass}\right)^2$ for the same 
reason that it breaks down in the 't Hooft model~\cite{thooft} and 
the spectrum becomes unbounded from below.  
Presumably this negative $\left(\mbox{mass}\right)^2$ term is
a signal for the presence of spontaneous symmetry breaking. 
One could imagine solving the broken phase of the theory using zero mode 
techniques~\cite{zeromode}. The resulting `shifted' theory would
have some complicated effective potential
representing the effects of the spontaneous symmetry breaking;
this effective potential is undoubtedly different than
the above form~(\ref{lac}).

\item 
Tadpole contractions associated with
four-point interactions, such as the first term in (\ref{lac2}), 
produce a divergent shift in the mass
term $\Tr \,\{M_i^\dd M_i\}$.  Renormalization leaves the 
mass as a free parameter in the Hamiltonian.

\item A successful renormalization group analysis would 
{\em introduce} couplings between neighboring links.
For instance, one expects terms associated with 
the constraint $2ag^2 M_i M_{i+1} \in SU(N)$.

\item From a more practical viewpoint, implementing the 
$\lambda_c \to \infty$ limit forces half of the dynamical degrees
of freedom at each link to decouple from the theory.  
This is not good news for numerical calculations where the 
computational difficulty depends critically on the number of 
dynamical degrees of freedom. 
\end{itemize}
As a consequence of these considerations, a first-principles construction
of the correct effective potential $V_i$ is not a simple matter.
Instead, we take a more pedestrian approach and include in $V_i$ 
all operators up to fourth order in $M_i$ containing one
color trace
\begin{eqnarray}
  V_{i} &=& \mu^2  \Tr \left\{M_{i}M_{i}^{\dd}\right\} 
+ {\lambda_1 \over a N}  \Tr \left\{M_{i}M_{i}^{\dd} 
   M_{i}M_{i}^{\dd} \right\} \nonumber \\
& &+ {\lambda_2 \over a N}  \Tr \left\{M_{i}M_{i+1} 
   M_{i+1}^{\dd}M_{i}^\dd\right\}
    \label{pot}
\end{eqnarray}
and try to determine the associated coupling constants empirically.
Note that the $\lambda_1$ term is local to one link and the $\lambda_2$
term acts on two adjacent links.

\subsection*{Quantization}

Next, we quantize the theory, write down the Hamiltonian $P^-$, and
construct a basis of states.  Further details may be found in 
Ref.~\cite{incest}.  At each lattice site $i$ we have a 1+1 dimensional
gauge theory with conserved current
\be
  J^\alpha_{i} = i \left( 
M_i \stackrel{\leftrightarrow}{\partial^\alpha} M_i^{\dd}+
M_{i-1}^{\dd} \stackrel{\leftrightarrow}{\partial^\alpha} M_{i-1} \right)
 \;.
\eq
We set $\partial_- A^+_i=0$ by choice of gauge and 
throw away the associated dynamical zero mode $\int dx^-  A^+_i$.  
The $A^{-}_i$ field obeys the 
equation of motion
\be
         \left(\partial_-\right)^2 A^{-}_i = \frac{g^2}{a} J^+_i 
	 \; . \label{aminus}
\eq
As evidenced by the absence of time derivatives $\partial_+$ in 
Eqn.~(\ref{aminus}), $A^-_i$ is not a true dynamical degree of
freedom but is constrained.  We solve the 
constraint equation (\ref{aminus}) 
for $A^-_i$  and remove it from the theory. 
At this point, quantization of the theory is straightforward.
The momentum conjugate to $M_i(x^-)$ is $\partial_- M_i^\dd(x^-)$ and
we impose the usual equal $x^+$ commutation relations\footnote{The factor
of 1/2 comes from the Dirac procedure for constrained systems. 
Also, we drop the constrained zero mode $\int dx^- M_i$.}
\be
       \left[M_i(x^-),\partial_- M_j^\dd(y^-)\right] = 
       		\frac{i}{2}\, \delta_{ij} \, \delta(x^- - y^-) \; .
\eq 
The longitudinal momentum operator
\be
  P^+ =  2 \sum_i  \int dx^-  \Tr  \left\{ \partial_- M_{i} 
\partial_- M_{i}^{\dd} \right\} \label{plus} 
\eq 
and Hamiltonian 
\be
 P^- = \sum_i  \int dx^- \left(- {g^2 \over 2a} \Tr \left\{ 
      J^{+}_{i} \frac{1}{\left(\partial_-\right)^2} J^{+}_{i}
           \right\} + V_{i} \right) \label{minus}
\eq
generate translations in the $x^-$ and $x^+$ directions, respectively.

Finally, we construct a basis of states. 
The zero mode of the $A^-_i$ constraint equation (\ref{aminus}) 
generates the Gau\ss' law constraint
\be
                 0 = \int dx^- J^{+}_{i} \label{gauss}
\eq
which we must impose on the basis of physical states.
That is, physical states must be color singlets at each lattice site.
We construct a basis of closed color loops:
\setlength{\unitlength}{60pt}
\be \begin{array}{rl}
\Tr\left\{ M_{i}M^{\dd}_{i}\right\} |0\rangle &
\raisebox{-.5em}{\begin{picture}(2.5,0.3)(-.25,-0.15)
\three\thicklines\put(0.5,0){\oval(1,.15)}
\end{picture}}\\[10pt]
\Tr\left\{ M_{i}M^{\dd}_{i} M_{i}M^{\dd}_{i}\right\} |0\rangle &
\raisebox{-.5em}{\begin{picture}(2.5,0.3)(-.25,-0.15)
\three\thicklines
\put(.5,0){\oval(1.1,.3)[l]}\put(.5,0){\oval(.9,.1)[l]}
\put(.5,0.1){\oval(1,.1)[r]}\put(.5,-.1){\oval(1,.1)[r]}
\end{picture}}\\[10pt]
\Tr\left\{ M_{i}M_{i+1} M_{i+1}^\dd M^{\dd}_{i}\right\} |0\rangle &
\raisebox{-.5em}{\begin{picture}(2.5,0.3)(-.25,-0.15)
\three\thicklines\put(1,0){\oval(2,.15)}
\end{picture}}\\[10pt]
\Tr\left\{M_{i}M_{i+1}M^{\dd}_{i+1}M_{i+1}M^{\dd}_{i+1}M_{i}^\dd
      \right\}|0\rangle&
\raisebox{-.5em}{\begin{picture}(2.5,0.30)(-.25,-0.15)
\three\thicklines
\put(1.5,0.1){\oval(1,.1)[r]}\put(1.5,-.1){\oval(1,.1)[r]}
\put(1.5,0){\oval(1,.1)[l]}\put(1.5,0){\oval(3,.3)[l]}
\end{picture}}\\[10pt]
\mbox{\em et cetera.}\hspace{0.5in}&
\begin{picture}(2.5,0.20)(-.25,0)
\put(0,0){\makebox[0in]{$i$}}\put(1,0){\makebox[0in]{$i+1$}}
\put(2,0){\makebox[0in]{$i+2$}}
\end{picture}
\end{array} \label{state}
\eq
where the $x^-$ co-ordinate of each link field remains arbitrary and the number
of links $M_{j}$ must equal the number of anti-links $M_{j}^\dd$. 
Since the loop-loop coupling constant is non-leading in $N$, 
we do not include states with more than one color trace in the Hilbert space; 
we deal with a free
string theory.\footnote{In fact it is possible to choose $V_i$ such that
one has a theory of free bosonic strings~\cite{suss}.} 
Incidentally, if we do introduce a state that is not a color singlet
at some site,  {\em exempli grati}
\be
\Tr\left\{ \cdots M_j M_{j+2} \cdots \right\} |0\rangle  \;
, 
\eq
we find that its energy diverges.

\subsection*{Confinement}

This theory has  built-in linear confinement.
Consider two test charges separated in the $x^-$ direction.
The first term in Eqn.~(\ref{minus}) acts as a linear confining
potential.\footnote{Note that 
$(\partial_-)^{-2}f(x^-) = \int dy^- |x^- - y^-| \, f(y^-)/2$
in Eqn.~(\ref{minus}).}
Now consider two charges at lattice sites $i$ and $j$. 
Due to the Gau\ss' law constraint (\ref{gauss})
we must construct a color string of at least $|i-j|$ link fields
between the two charges.
If $\mu^2>0$, there is some minimum energy associated
with each link field and the transverse 
string tension is nonzero.  As we shall see in the next section
the resulting potential is, in fact, linear.
Similar arguments hold for the original action (\ref{lag}).

\subsection*{Numerical Techniques}

We solve the spectrum on the computer using DLCQ techniques~\cite{brod}.
For the $x^-$ coordinate, we impose anti-periodic boundary conditions 
$M_i(x^-) = - M_i(x^-+L)$ and use a momentum space 
representation.  For integer valued cut-off $K= L P^+/(2\pi)$, 
momenta are labeled by odd half integers 
$\kappa_m \in \{1/2,3/2,\ldots\}$ where $\sum_m \kappa_m = K$.
This yields a finite basis of states. 
Also, we employ a truncation in particle number.

The first term of the Hamiltonian (\ref{minus}) dominates the behavior
of the theory.  Thus it is natural to choose units such
that the associated coupling constant $g^2 N/a$ is set
equal to $1$. This choice of units is assumed in
the following.

\section*{STRING TENSION}

Next, we introduce a method for measuring the 
string tension in the $x^1$ direction. 
Consider a lattice of $n$ transverse links and periodic
boundary conditions.  We construct a basis of states that 
wind once around this lattice
and calculate the lowest eigenvalue of the invariant 
$(\mbox{mass})^2$ operator $M^2 |\Psi\rangle = 2 P^+ P^- |\Psi\rangle$.
\begin{figure}
\centering
\BoxedEPSF{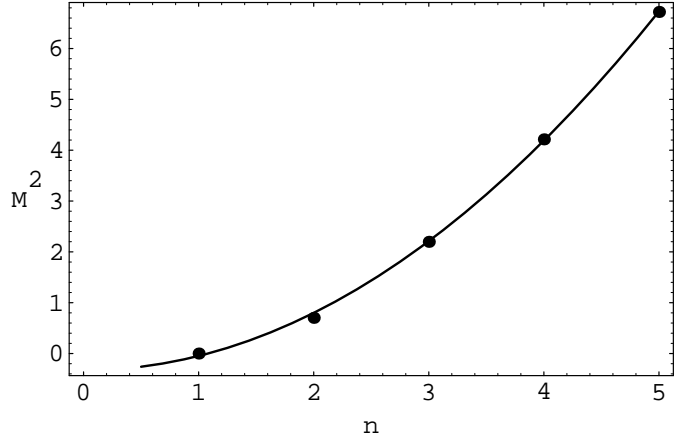 scaled 1102}
\caption{Lowest $M^2$ eigenvalue vs 
$n$ for states that wind once around the lattice.
Here, $K=10.5$ or 11 (($n+4$)-particle truncation),
$\mu^2=0.1$, $\lambda_1/a =1$, and $\lambda_2/a=-1$.
Also shown is a fit to a quadratic. \label{fig1}}
\end{figure}
The continuum limit string tension is
\be
 \lim_{a \to 0} \frac{1}{a}\frac{\Delta M}{\Delta n} \;, \;\;\;\; 
      \mbox{$n a$ fixed.}
\eq
This definition is be equivalent to the standard definition of 
string tension using two test charges if the charges are placed 
sufficiently far apart.
If we plot $M^2$ vs.\ $n$ in Figure~\ref{fig1}, 
we obtain a quadratic corresponding to a linear confining 
potential. 

\begin{figure}
\centering
\BoxedEPSF{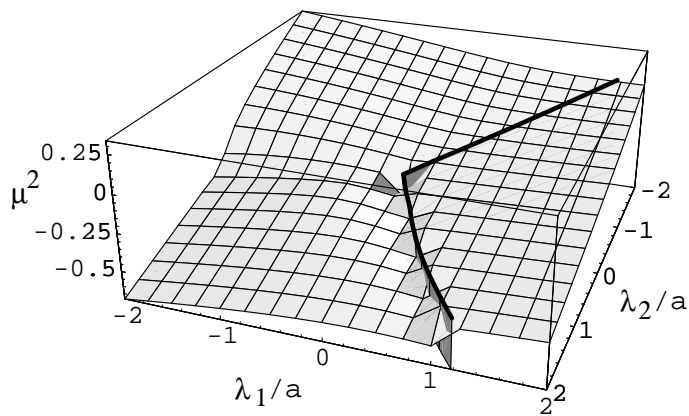 scaled 1102}
\caption{ 
Parameters such that the lowest $M^2$
eigenvalues are equal for $n=4$ and $5$ (see Fig.~\protect\ref{fig1}), 
where $K=10.5$ or $11$ (($n+4$)-particle truncation).  
This is an estimate of vanishing string
tension.  Also shown is a line such that the $M^2$ eigenvalues are 
approximately degenerate for $n=3$, $4$, and $5$. \label{fig2}}
\end{figure}

In fact, we find that the string tension decreases with decreasing
$\mu^2$, leading us to the conclusion that $\mu^2$ is a measure 
of the physical lattice spacing. 
The continuum limit is partially fixed by requiring
vanishing $\Delta (M)/\Delta (n)$; see Fig.~\ref{fig2}.
Above the surface, the string tension is positive corresponding
to nonzero lattice spacing.

In addition we have an analytic ansatz for these states
that wind once around the lattice
that is valid in the $\lambda_2/a<0$ region of parameter space.
It agrees well with the numerical results.

\section*{SPECTRUM}

\begin{figure}
\centering
\BoxedEPSF{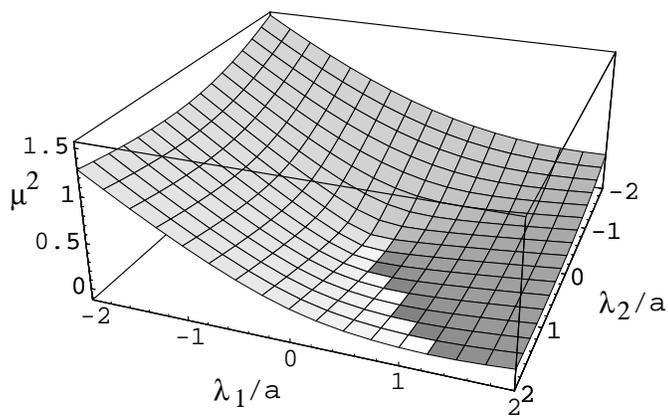 scaled 1102}
\caption{
Parameters such that the lowest $M^2$
eigenvalue is zero, $K=10$ to $14$ 
with extrapolation using a fit to
$\sum_{m=0}^{4} c_m K^{-m/2}$ (6-particle truncation).  
Below this surface, the spectrum
is tachyonic; above the surface, it is well behaved. \label{fig3}}
\end{figure}

Before we look at the details of the spectrum,
we can make some statements about the allowed region
in parameter space.  A numerical estimate
of the `edge' of this region is plotted in Fig.~\ref{fig3}.  
Comparing Figs.~\ref{fig2} and 
\ref{fig3},
we see that a continuum limit with a non-tachyonic spectrum occurs only for
the ``wedge shaped region'' $-\lambda_1 \leq \lambda_2 \leq \lambda_1/2$.
In this region, the continuum limit occurs for
$\mu^2 \approx 0$ to within numerical errors.
Finite lattice spacing corresponds to $\mu^2$ slightly above the
surface in Fig.~\ref{fig2}.

\subsection*{Symmetries}

The theory possesses several discrete symmetries. Charge conjugation
induces the symmetry ${\cal C}: \, (M_i)_{l,m} \leftrightarrow 
(M_i^\dd)_{m,l}$ where $l,m\in \{1,\ldots,N\}$. 
Parity is the product of two reflections
${\cal P}_1: \, x^1 \to -x^1$
and ${\cal P}_2: \, x^2 \to -x^2$.
In light-front quantization, ${\cal P}_1$ is an exact symmetry 
${\cal P}_1:  \, M_i \leftrightarrow M_{-i}^\dd$ while
${\cal P}_2$: $x^+ \leftrightarrow x^-$, is complicated. 
Its explicit operation is known only for free
particles \cite{horn}, which we call ``Hornbostel parity.''\@  
The latter is
nevertheless useful since it is often an approximate quantum number
and can be used to estimate ${\cal P}_2$~\cite{tube}.
Given ${\cal P}_2$ and ${\cal P}_1$ we can determine whether spin
${\cal J}$ is even or odd using the relation 
$(-1)^{{\cal J}} = {\cal P}_1 {\cal P}_2$.
If rotational symmetry has been restored in the theory, 
states of spin ${\cal J} \neq 0$ should 
form degenerate ${\cal P}_1$ doublets 
$|+{\cal J}\rangle \pm |-{\cal J}\rangle$~\cite{teper}.
As with the lattice results, we use ``spectroscopic notation'' 
$|{\cal J}|^{{\cal P}_1 {\cal C}}$ to classify states.

One expects the lowest
two eigenstates to be approximately two-particle states
\be
 \sum_i \Tr\left\{M_i^\dd(x^-) M_i(y^-) \right\} |0\rangle
\eq
with the lowest state having a symmetric wavefunction
corresponding to $0^{++}$ 
and the first excited state having an antisymmetric wavefunction 
corresponding to $0^{--}$. 
Of course, these states also have 6-particle {\em et cetera}
contributions.

\subsection*{Results}

\begin{figure}
\centering
\BoxedEPSF{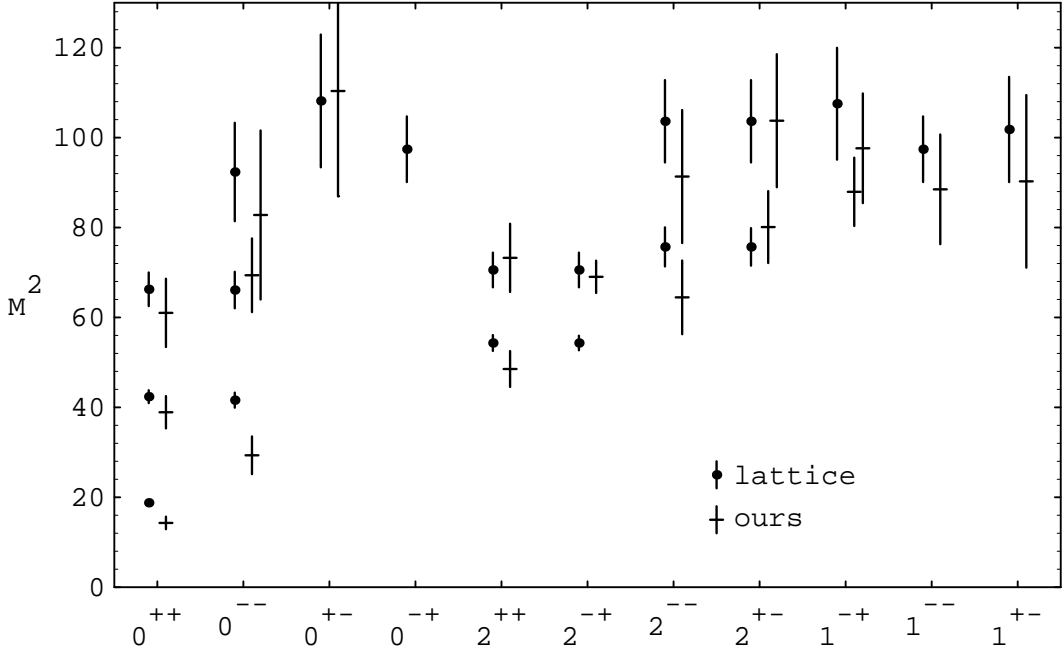 scaled 1098}
\caption{
A comparison of our spectrum with $SU(3)$ ELMC data in units of the 
physical string tension \protect\cite{teper} 
for various $|{\cal J}|^{{\cal P}_1 {\cal C}}$.  
The parameters $g^2 N/a=3.90$, $\mu^2=0.134 g^2 N/a$,
$\lambda_1 = 0.487 g^2 N$, and $\lambda_2 = 1.108 g^2 N$ were chosen by
a best fit to the lattice data, $\chi^2 = 40$, where $K=10$ 
(8-particle truncation).
Our error estimates are solely for the 
purpose of performing the $\chi^2$ fit.  
\label{fig4}}
\end{figure}
Ideally, we would like to predict the effective
potential based on some connection to the continuum theory,
restoration of rotational invariance, {\em et cetera}.
However, as a first step, we use instead a best $\chi^2$ 
fit to the ELMC results of Teper \cite{teper}.

An example spectrum is shown in Fig.~\ref{fig4}.
Similar spectra are found in other regions of coupling
constant space above the `wedge-shaped region' 
(Fig.~\ref{fig3}).
We label the lowest $2^{--}$ and second $0^{--}$ states based on the
expectation value of the number operator and determine  
$(-1)^{\cal J}$ based on Hornbostel parity \protect\cite{tube};  
the exception is the $\left|{\cal J}\right|^{+-}$ sector where
Hornbostel parity gave exactly the opposite of the desired results.
Beyond this, $J$ is determined by a best fit to the lattice data.

Since this spectrum is the result of a best fit, it is
not very predictive.  However, we can use the result to
tell us about our model.  At first glance, we seem to 
have a pretty good match.  However, we note several problems:
\begin{itemize}
\item  The energy of the lowest $0^{--}$ state is too low. 
\item  The lowest $2^{++}$ and $2^{-+}$ states form a degenerate
       doublet if rotational symmetry is restored (as indeed
       happens for the lattice data).  In our case,
       the splitting is large. This discrepancy dominates
       the error in our $\chi^2$ fitting procedure.
\end{itemize}
Let us review the possible sources of error in our calculation.
\begin{itemize}
\item[] {\bf Large N.}  We compare $N\to \infty$ spectra
     to $SU(3)$ lattice results.  However, based on 
     lattice calculations for $SU(2)$, $SU(3)$, and $SU(4)$,
     $1/N$ corrections 
     to the low energy spectrum are   small~\cite{teper}.
     
\item[] {\bf Finite K.} Our discretization of the longitudinal
     momentum introduces some error.  However, we have generated
     spectra for $K=10,11,12,13,14$, extrapolated
     to large $K$ (6-particle truncation),
     and compared to large $N$ extrapolated ELMC
     spectra.  We saw no real improvement in our results.

\item[] {\bf Particle Number.}  We also impose a truncation in the
     number of particles.  We have examined spectra for 
     4-, 6- , and 8-particle truncations ($K=10$), extrapolated to 
     large number of particles,
     and compared to large $N$ extrapolated ELMC
     spectra.  We saw no real improvement in our results.

\item[] {\bf Hamiltonian.} The effective potential $V_i$ that we
     chose (\ref{pot}) did not contain any 6-point or higher
     interactions.  In addition, we did not include any operators
     containing multiple traces, for instance
     $(\Tr \, \{M_i M_i^\dd\})^2$.\footnote{Although this term is generally
     not leading order in $N$, it does act on the two particle 
     subspace of the theory.}

\end{itemize}

\section*{CONCLUSIONS}

We have investigated the transverse lattice model of
Bardeen and Pearson \cite{bard1} for $(2+1)$-dimensions in the large-$N$ 
limit using linearized link variables and an empirical effective 
potential $V_i$. We identified a choice for $V_i$ corresponding to 
vanishing string tension. The glueball spectrum
in the vicinity agreed qualitatively with that coming from 
the presumably reliable Euclidean lattice Monte Carlo 
results~\cite{teper}. Most importantly
however, we did not see significant signs of rotational invariance
which could lead one to conclude that the transverse gauge dynamics were
correctly accounted for by $V_i$.  We believe that our choice 
(\ref{pot})
is probably too simple and that higher order terms are necessary to see
improvement in our spectrum. 

Future work includes the addition of more operators in the 
effective potential.  Also, we can use our method of measuring 
string tension to measure the physical lattice spacing.  
This issue needs further investigation.
Most importantly, we need a more concrete connection 
between our model and the continuum theory.  
This would allow us to better predict the
correct effective potential.

\section*{ACKNOWLEDGMENTS}

This work was supported, in part, by the 
Alexander Von Humboldt Stiftung. B. van de Sande would
like to thank the organizers of {\sc Orbis Scientiae 1996}
for providing a stimulating and enjoyable atmosphere.


\end{document}